# Magnetocaloric Effect of Gadolinium at Adiabatic and Quasi-Isothermal Conditions in High Magnetic Fields

Alexander P. Kamantsev[1,2,a*], Victor V. Koledov[1,2], Alexey V. Mashirov[1,2], Elvina T. Dilmieva[1], Vladimir G. Shavrov[1], Jacek Cwik[2], Irina S. Tereshina[2,3]

[1]Kotelnikov Institute of Radio-engineering and Electronics of RAS, Moscow, 125009, Russia
[2]International Laboratory of High Magnetic Fields and Low Temperatures, Wroclaw, 53-421, Poland
[3]Baikov Institute of Metallurgy and Material Science of RAS, Moscow, 119991, Russia
[a]kama@cplire.ru

**Abstract.** High cooling power of magnetocaloric refrigeration can be achieved only at large amounts of heat, which can be transferred in one cycle from cold end to hot end at quasi-isothermal conditions. The simple experimental method for direct measurement of the transferred heat from material with magnetocaloric effect (MCE) to massive nonmagnetic block at quasi-isothermal conditions was proposed. The vacuum calorimeter was designed for the simultaneous measurements of MCE both at adiabatic conditions ($\Delta T$) and quasi-isothermal conditions ($\Delta Q$) in the magnetic fields of Bitter coil magnet. This calorimeter was tested on samples of pure polycrystalline Gd with direct MCE. The maximal obtained values were $\Delta T$ = 17.7 K and $\Delta Q$ = 5900 J/kg at initial temperature 20 $^0$C in magnetic field 140 kOe.

## Introduction

A big interest is attracted to the application of materials with a large MCE in vicinity of magnetic and magnetostructural phase transitions (PT) for creation of household and industrial refrigerators, operating at room temperature. High cooling power of such devices can be achieved only at high frequency of heat transfer cycles and at large amounts of heat, which can be transferred in one cycle from cold end to hot end. The maximal frequency of cycles depends on the geometry of the working body made of material with MCE and the fundamental restrictions on the rate of PT in this material. The optimum configuration of working body is expected to be the plates in the honeycomb structure [1]. The rate of PT depends on the relaxation processes in the vicinity of PT. The magnetization relaxation time of the classical magnetocaloric material Gd near the Curie point is discussed in terms of Landau-Khalatnikov equation, and the value $\tau \approx$ 50 ms was obtained experimentally in [2]. It restricts the possible operating frequency of cycles of magnetocaloric refrigeration based on Gd working body. In the present work we concentrate on experimental measurement of transferred heat in Gd. We suggest the new experimental approach for the direct measurements of MCEs at quasi-isothermal $\Delta Q$ and adiabatic conditions $\Delta T$ simultaneously. We report the results of direct measurements of $\Delta T$ and $\Delta Q$ in high magnetic fields of Bitter coil magnet up to 140 kOe. On the base of the results of these measurements we estimate the possible specific cooling power of the working body of the hypothetic magnetocaloric refrigerator made of Gd plates.

## Theory

If a sample of magnetocaloric material is placed in good thermal contact with a massive nonmagnetic block with determined specific heat and high thermal conductivity (see Fig. 1), then we can experimentally estimate the transferred heat (per unit of the sample mass) $\Delta Q$ from the sample to the block at quasi-isothermal conditions by measuring $\Delta T_b$ – the quasi-isothermal temperature change of the block at magnetic field change

$$\Delta Q \times m = M_b \times C \times \Delta T_b + m \times c \times \Delta T_b, \qquad (1)$$

where $M_b$ – the mass of the block, $C$ – the specific heat of the block, $m$ – the mass of the sample, $c$ – the specific heat of the sample. We can neglect the heat, which connected with the temperature change of the sample, if the mass of the sample is negligible compared to the mass of the block ($M_b \gg m$), then

$$\Delta Q \approx (M_b/m) \times C \times \Delta T_b. \qquad (2)$$

If the working body has a cellular structure consisting of very thin plates of magnetic material [1] and the frequency of heat transfer cycles is limited to the relaxation time of magnetization near PT [2], we can calculate the achievable specific cooling power of the working body of the magnetic refrigerator as

$$P = \Delta Q \times f, \qquad (3)$$

where $f = 1/(2\tau)$ is the maximal possible frequency of heat transfer cycles.

**Experiment**

The experimental vacuum calorimeter for simultaneous $\Delta T$ and $\Delta Q$ measurements in high magnetic fields was developed. The measured samples of pure Gd were: bulk "Sample_1" and "Sample_2" in the shape of plate. The samples had the Curie temperature $T_C = 20\ ^0C$ and direct MCE. The Sample_1 ($M = 4.779$ g) was placed into vacuum chamber at adiabatic conditions. The Sample_2 ($m = 0.565$ g) was glued (by heat-conducting glue) on massive copper block ($M_b = 7.404$ g) for providing the quasi-isothermal conditions and was placed into the same vacuum chamber (see Fig 1). The appearance of the textolite framework with the samples is shown on Fig.2. The temperature measurements were performed with the help of semiconductor light diodes (LED) under constant current (100 µA). The temperature sensitivity of these diodes is constant in wide region near the room temperature and equals to 1.0 mV/K. The magnetic field does not affect the sensitivity of such diodes. The mass of one the diode is 1.5 mg.

The protocol of experiment was as follows. The vacuum calorimeter was placed into the water/ice thermostat in Bitter coil magnet. The vacuum in the chamber was achieved by forepump (down to pressure 0.4 Pa). After that, the temperature measurements started. The initial temperature was adjusted with the help of thermo-control system. The magnetic field was turned on with the constant rate 2.0 kOe/sec. The magnitude of magnetic field was measured by Hall probe. The similar experimental technique was applied for metamagnetic $Ni_{43}Mn_{37.9}In_{12.1}Co_7$ Heusler alloy in [3,4] ($\Delta T$ and $\Delta Q$ were measured nonsimultaneously).

The extraction technique was applied for increasing the accuracy of $\Delta T$ and $\Delta Q$ measurements. The thermostat with the vacuum chamber was extracted from Bitter coil magnet with the help of special mechanical system before turning on the magnetic field. Next we established a required magnetic field 0 – 140 kOe. Then the thermostat with the samples in vacuum chamber could be quickly inserted into and extracted from the region of magnetic field (with the maximal rate ±50 kOe/sec). This method allows to improve the adiabatic conditions in a vacuum chamber and to increase the accuracy of the experiments.

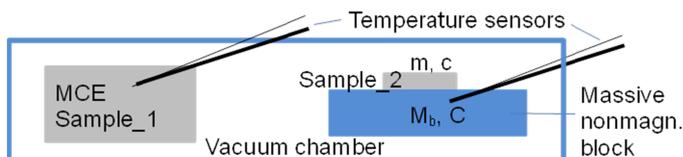 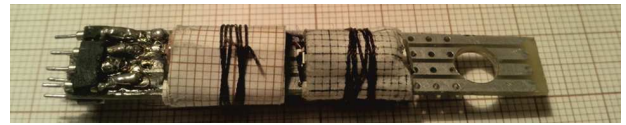

Fig.1. The scheme of the experimental setup for direct $\Delta T$ and $\Delta Q$ measurements simultaneously in Bitter coil magnet.

Fig.2. The appearance of the textolite framework with Gd samples for direct $\Delta T$ and $\Delta Q$ measurements in Bitter coil magnet.

**Results**

The typical experimental data are presented on Fig.3. The maximal obtained value of adiabatic MCE was ΔT = 17.7 K at initial temperature $T_0 = 20\ ^0C$ in magnetic field H = 140 kOe on Sample_1. The transferred heat ΔQ from Sample_2 to copper block was calculated using Eq. (2) and presented on Fig. 4. The maximal value of ΔQ = 5900 J/kg at $T_0 = 20\ ^0C$ in H = 140 kOe was found. These data were obtained with using the extraction technique, without extraction the values of MCE were less on 10-15 %.

Our results for ΔT in Gd (Fig. 5) have good correlations with data obtained elsewhere [2,5]. One can compare our results for ΔQ with the data obtained by indirect methods. For example, we have ΔQ = 1300 J/kg at $T_0 = 20\ ^0C$ in H = 20 kOe (Fig. 6). There is ΔQ = 1600 J/kg, calculated from heat-capacity and magnetization measurements for Gd single crystal in the same conditions [5].

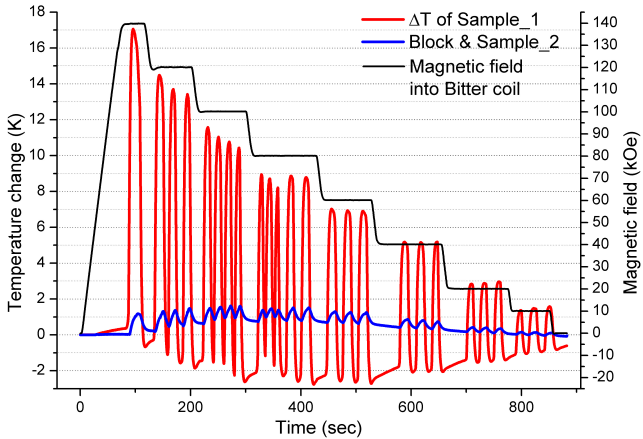

Fig.3. The time dependence of magnetic field change and corresponding ΔT of Sample_1 and $\Delta T_b$ of the copper block with Sample_2 with using of extraction technique. $T_0 = 20\ ^0C$.

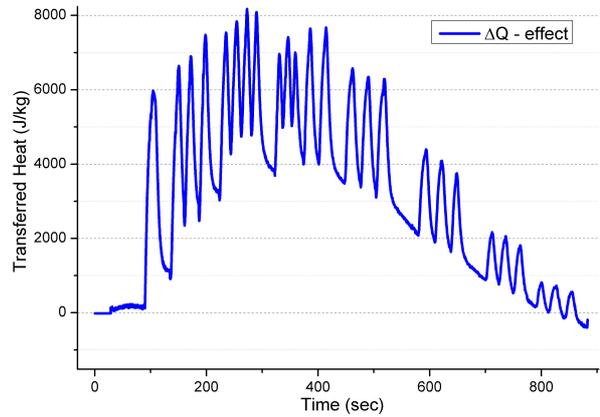

Fig.4. The transferred heat ΔQ from the Gd Sample_2 to the copper block at quasi-isothermal conditions. It was calculated from data on Fig.3 using Eq. (2).

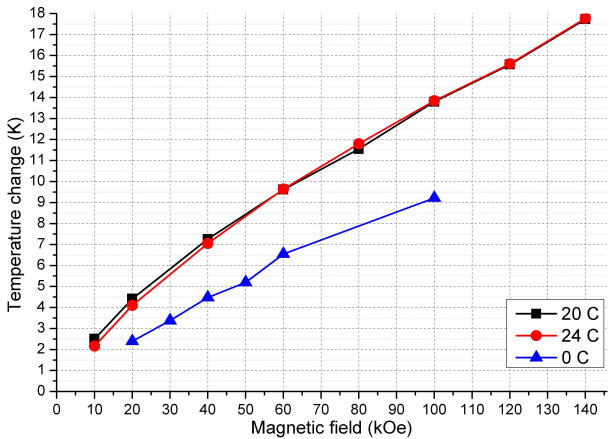

Fig. 5. The adiabatic temperature change ΔT of Gd Sample_1 as a function of magnetic field at different initial temperatures.

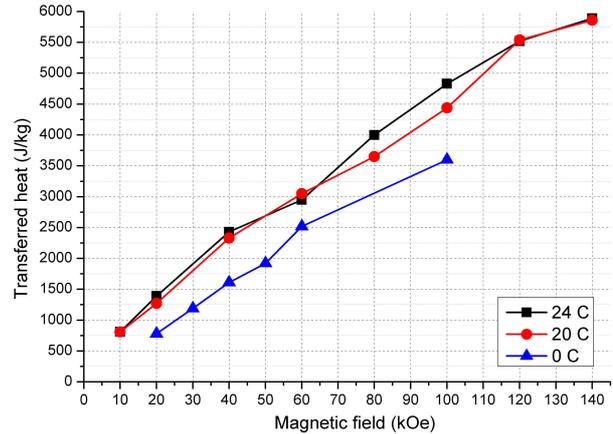

Fig.6. The transferred heat ΔQ from Gd Sample_2 to copper block as a function of magnetic field at different temperatures.

Let us estimate the achievable specific cooling power of the working body of the magnetic refrigerator working near room temperature. The magnetization relaxation time is τ = 50 ms at $T_0 = 20\ ^0C$ [2]. Therefore, the frequency of the cycles of a magnetic refrigerator with a Gd working body will be restricted to f = 1/(2τ) = 10 Hz. Then, with the help of Eq. (3) we get the following estimation in H = 20 kOe: P = 1300 J/kg × 10 Hz = 13 W/g. The maximal value of specific cooling power of Gd working body will be: P = 59 W/g in H = 140 kOe, if the relaxation time does not depend on magnetic field.

**Conclusions**

The MCE in Gd was measured in direct experiments at adiabatic ΔT and quasi-isothermal ΔQ conditions simultaneously in high magnetic fields up to 140 kOe. The maximal obtained values were ΔT = 17.7 K and ΔQ = 5900 J/kg at $T_0$ = 20 $^0$C in H = 140 kOe. The MCE measured at quasi-isothermal conditions (the transferred heat) ΔQ is the more practical quantity than the magnetic entropy change, because it makes easy possible the robust estimation of the cooling power of hypothetic magnetocaloric refrigerator.

The extraction technique improves the experimental conditions and increases the accuracy of direct experiments on 10-15 %.

The investigation of possibility of the high frequency of heat exchange cycles f > 10-100 Hz is crucial for a design of magnetic refrigerator. Following our experimental data, the achievable cooling power (per one gram of material) of the hypothetic magnetic refrigerator based on working body made of Gd plates can be estimated as: P = 13 W/g in H = 20 kOe and P = 59 W/g in H = 140 kOe at $T_0$ = 20 $^0$C.

**Acknowledgments**

The reported study was supported by Russian Science Foundation, grant No. 14-22-00279.